# Unexpected Robustness of the Band Gaps of TiO$_2$ under High Pressures

Yong Yang

*Key Laboratory of Materials Physics, Institute of Solid State Physics, Chinese Academy of Sciences, Hefei 230031, China.*

**ABSTRACT:** Titanium dioxide (TiO$_2$) is a wide band gap semiconducting material which is promising for photocatalysis. Here we present first-principles calculations to study the pressure dependence of structural and electronic properties of two TiO$_2$ phases: the cotunnite-type and the Fe$_2$P-type structure. The band gaps are calculated using density functional theory (DFT) with the generalized gradient approximation (GGA), as well as the many-body perturbation theory with the *GW* approximation. The band gaps of both phases are found to be unexpectedly robust across a broad range pressures. The corresponding pressure coefficients are significantly smaller than that of diamond and silicon carbide (SiC), whose pressure coefficient is the smallest value ever measured by experiment. The robustness originates from the synchronous change of valence band maximum (VBM) and conduction band minimum (CBM) with nearly identical rates of changes. A step-like jump of band gaps around the phase transition pressure point is expected and understood in light of the difference in crystal structures.

# 1. Introduction

Pressure is a powerful tool for tuning the atomic and electronic structures of materials. Both inorganic and organic substances, such as Si, Ge, $H_2O$, $SiO_2$ and $C_6H_6$, can exhibit complex condensed phases under high pressures [1-5]. With the application of high pressure, people are able to synthesize some novel compounds that do not exist at ambient conditions [6]. Tremendous studies have demonstrated that the electronic properties of materials can be effectively manipulated by compression, through which the metal-to-semiconductor [7, 8] or insulator-to-metal transitions [9-15] take place. It is conjectured that insulators and semiconductors always metallize under sufficiently high external pressures [16]. The most notable examples may be the prediction of possible metallization and superconductivity in solid state hydrogen [17-19], and the observation of high-$T_c$ superconductivity in hydrogen sulfide under high pressures [20, 21].

In principle, the electronic structures of materials under high pressures can be probed by experimental measurements on the quantities such as electrical conductivity, optical conductivity and reflectivity [22, 23]. To date, however, measurements on the electronic properties of wide band gap semiconductors are limited to pressures of several to tens GPa. At extreme pressures below the pressure point of metallization (rarely attainable in experiment), typically ~ 100 to 200 GPa, little is known about the characteristics of electronic structures of the *intermediate states* of wide band gap semiconductors. This may be due to the technique difficulties in accurate determination of electronic band structures at the presence of high pressure devices like the diamond anvil cell (DAC), or the intrinsic or impurity absorption edges of diamond in conventional optical spectroscopy [22]. There lacks the knowledge about the global behavior of electronic structures across a wide range of pressures, in particular, pressures spanning an interval with the width of ~ 100 GPa or larger.

In this work, we study the effects of pressure on titania ($TiO_2$), one of the most extensively studied wide band gap semiconductors [24-26], at pressures spanning over an interval of ~ 150 GPa. As an important photocatalyst for harvesting solar energy, the alignments of energy levels near the valence and conduction bands play a key role [25, 27]. At ambient pressure, TiO$_2$ commonly exist in natural minerals in three phases: rutile, anatase, and brookite [25]. All the three phases have an optical band gap of slightly higher than 3 eV [28-30]. At elevated pressures, TiO$_2$ undergoes phase transformations from the most stable rutile structure at ambient pressure to a number of high-pressure polymorphs [31, 32]. For pressures above 50 GPa, TiO$_2$ adopts an orthorhombic cotunnite-type structure [32] until ~ 160 GPa, at which a new phase with a hexagonal Fe$_2$P-type structure will prevail [33]. At ~ 650 GPa, recent theoretical studies have predicted the existence of a new tetragonal phase (space group: $I4/mmm$), in which the metallization of TiO$_2$ may occur [34, 35]. The present work is focused on the pressure dependence of electronic structures of two experimentally verified high-pressure TiO$_2$ phases: the cotunnite-type [31, 32] and the Fe$_2$P-type structure [33]. At moderate or low pressures, there have been a lot of researches on the pressure dependence of the band gaps of diamond-, zinc-blende-, and wurtzite-structure semiconductors [36-39]. These works show that, the band gap ($E_g$) can be expressed as a function of pressure ($P$) [37-39]:

$$E_g(P) = E_g(0) + \alpha P + \beta P^2, \qquad (1)$$

where $\alpha$ and $\beta$ are the parameters to be determined, and $c_p = dE_g/dP$ is the so-called *pressure coefficient* [36-39], which is a key parameter describing the pressure dependence of band gaps. It is *unclear* whether Eq. (1) would be applicable or not to wide band gap semiconductors at $P \geq 100$ GPa. From the data reported in literatures [36-50], previous studies are focused on the pressure coefficients ($dE_g/dP$) at ambient or moderate pressures, at which the values of $c_p$ can be either positive or negative, with an order of magnitude of 10$^{-2}$ to 10$^{-1}$ eV/GPa for most systems. By contrast, our studies based on first-principles calculations find that, the band gaps of the two high-pressure phases of TiO$_2$ are surprisingly *robust* to compression. Within a wide range of pressures (50 to 200 GPa for the cotunnite-type, and 110 to 260 GPa for the Fe$_2$P-type), the pressure coefficients of TiO$_2$ are significantly smaller than that of diamond (~ 5 meV/GPa) [41, 42] and silicon carbide (SiC) (~ 1.9 meV/GPa for the cubic SiC [43], and ~ 2 meV/GPa for 6$H$-SiC [44]), the smallest value ever measured experimentally. Furthermore, we find that the pressure dependence of the band gaps of both phases is well described by a modified form of Eq. (1).

## 2. Theoretical methods

The first-principles calculations were carried out by the Vienna *ab initio* simulation package (VASP) [51, 52], which is based on density functional theory (DFT). A plane wave basis set and the projector-augmented-wave (PAW) potentials [53, 54] are employed to describe the electron-ion interactions. The exchange-correlation interactions of electrons are described by the generalized gradient approximation (GGA) within the PBE formalism [55]. The energy cutoff for plane waves is 600 eV. For the structural optimization and total energy calculations, a 6×10×5 and 6×6×10 Monkhorst-Pack k-mesh [56] is generated for sampling the Brillouin zone (BZ) of the cotunnite-type and $Fe_2P$-type structure, respectively. These set of parameters yield convergence of total energies to within a level of $10^{-2}$ meV/$Ta_2O_5$ unit. Since DFT is a ground state theory and not suitable to estimate the band gap values, we have employed the *GW* method [57, 58], which explicitly includes the many-body effects (exchange and correlation) of electrons to calculated the energy levels of low-lying excited states and consequently the band gaps of $TiO_2$ under high pressures. Specially, we take the one-shot *GW* approach implemented in the VASP code [59] to calculate the energy spectrum. The quasiparticle energies and wave functions are obtained by solving a Schrödinger-type equation [58]:

$$(T + V_{ext} + V_H)\psi_{nk}(\vec{r}) + \int d\vec{r}\,'\Sigma(\vec{r},\vec{r}';E_{nk})\psi_{nk}(\vec{r}') = E_{nk}\psi_{nk}(\vec{r}), \qquad (2)$$

where $T$ is the kinetic energy operator of electrons, $V_{ext}$ is the external potential due to ions, $V_H$ is the electrostatic Hartree potential, $\Sigma$ is the electron self-energy operator, and $E_{nk}$ and $\psi_{nk}(\vec{r})$ are the quasiparticle energies and wave functions, respectively. Within the *GW* approximation proposed by Hedin [57], the term $\Sigma$ can be calculated as follows:

$$\Sigma(\vec{r},\vec{r}';E) = \frac{i}{2\pi}\int d\omega e^{i\delta\omega}G(\vec{r},\vec{r}';E+\omega)W(\vec{r},\vec{r}';\omega), \qquad (3)$$

where *G* is the Green's function, *W* is the dynamically screened Coulomb interaction,

and δ is a positive infinitesimal. In the *GW* calculations on the cotunnite-type and $Fe_2P$-type $TiO_2$, a 4×6×3, 4×4×6 k-mesh is used to sample the BZ, respectively. The number of energy bands involved in the *GW* calculations is 192 for both systems.

**3. Results and discussion**

*3.1. Lattice Parameters and Enthalpy as a Function of Pressure*

The dependence of the lattice parameters with pressure is shown in Figures 1(a)-(b), for the cotunnite-type and $Fe_2P$-type $TiO_2$. For both phases, the lengths of cell axes decrease almost linearly with pressure. Rate of compression is $2.34 \times 10^{-3}$ Å/GPa in *a*-axis, $1.42 \times 10^{-3}$ Å/GPa in *b*-axis, $2.19 \times 10^{-3}$ Å/GPa in *c*-axis for cotunnite-type $TiO_2$; and $1.47 \times 10^{-3}$ Å/GPa in *a*-axis (and *b*-axis), $1.15 \times 10^{-3}$ Å/GPa in *c*-axis for $Fe_2P$-type $TiO_2$. Due to their large bulk moduli [32, 33, 60], both phases show considerable incompressibility. The calculated enthalpy (*H*) as a function of pressure is shown in Figure 1(c). The two lines joining the data points almost superpose each other, despite that the $Fe_2P$-type structure is energetically slightly favored over the cotunnite-type structure at ~ 115 GPa and above (see the inset of Figure 1(c), for $\Delta H$). The predicted phase transition pressure (~ 115 GPa) is ~ 45 GPa smaller than previously reported (~ 160 GPa for DFT-GGA calculations) [33], which may arise from the difference of simulation parameters [see Ref. 61 for details].

On the other hand, the coincidence of the two enthalpy lines suggests the *coexistence* of both phases across a broad range of pressures. At *T* = 300 K, one can estimate the ratio of probability of finding the two phases ($Fe_2P$-type versus cotunnite-type): $r = \exp[-\Delta H/(k_B T)]$, which turns out to be 0.946, 1.066, 1.127, 1.826, 2.252, 3.400, for *P* = 110, 115, 120, 150, 160, and 200 GPa, respectively. Indeed, previous experimental work has shown that the cotunnite-type structure preserves at ambient pressure upon rapid decompression and quenched in liquid nitrogen (77 K) [62]. Moreover, the coexistence of different $TiO_2$ phases below 60 GPa has also been experimentally demonstrated by compression-decompression method using the diamond-anvil cell (DAC) technique [32]. In this context, the coexistence of the two high-pressure phases studied here is naturally expected for *P* > 100 GPa.

### 3.2. Band Gaps as a Function of Pressure

The calculated band gaps of the cotunnite-type and Fe$_2$P-type TiO$_2$ are shown in Figures 2(a) and 2(b), as a function of applied pressure. The band gap values obtained by using traditional DFT-GGA calculations are shown along with those obtained using the *GW* method. Despite some minor fluctuations, the band gaps of both high-pressure phases of TiO$_2$ are rather robust to compression. To get a deep understanding on this phenomenon, the data are least-squares fitted to a modified form of Eq. (1) as follows:

$$E_g(P) = E_g(P_0) + \alpha(P - P_0) + \beta(P - P_0)^2, \qquad (4)$$

where $\alpha$ and $\beta$ have the same meanings as in Eq. (1); $E_g(P_0)$ is the band gap at the starting point of pressure range under investigation. The value of $P_0$ is 50 GPa and 110 GPa, respectively, for the cotunnite-type and Fe$_2$P-type TiO$_2$. The fitting parameters are summarized in Table I. Consequently, the pressure coefficient is calculated as the first derivative of $E_g$ with respect to $P$:

$$c_p = \alpha + 2\beta (P - P_0). \qquad (5)$$

This is a linear function of pressure $P$. Specially, $c_p = \alpha$ when $P = P_0$. The pressure variation of $c_p$ is shown in Figures 2(c) and 2(d), for the two phases. In the case of GGA calculations, the value of $\alpha$ is 0.117 meV/GPa and -0.813 meV/GPa, for the cotunnite-type and Fe$_2$P-type TiO$_2$, respectively. For the more accurate *GW* method, $\alpha$ is 1.831 meV/GPa for the cotunnite-type and 0.489 meV/GPa for the Fe$_2$P-type structure. As seen from Table I, all the values of $\beta$ are the order of magnitude of $-10^{-3}$ meV/GPa$^2$, which result in the linear decrease of $c_p$ with pressure. Meanwhile, the point $c_p = 0$ gives the pressure $P_m = P_0 - \alpha/(2\beta)$, at which the band gap $E_g$ reaches its maximum. For instance, from the parameters describing the *GW* band gaps (Table I), $P_m$ is calculated to be ~ 180.1 GPa and ~ 140.0 GPa, for the cotunnite-type and Fe$_2$P-type structure, respectively. Compared to the pressure coefficient of diamond (~ 5 meV/GPa) [41, 42] and that of SiC (~ 1.9 meV/GPa for cubic SiC and ~ 2 meV/GPa for 6*H*-SiC) [43, 44], which is the smallest value ever measured by experiments, the pressure coefficients of the two high-pressure TiO$_2$ phases found in our work are still significantly smaller. In particular, the region in which $|c_p| \leq 0.5$ meV/GPa, i.e., one order of magnitude smaller than that of diamond, is highlighted in Figures 2(c) and 2(d). For the data obtained by *GW* calculations, the highlighted region spans a pressure interval of [144.6 GPa, 215.7 GPa] for the cotunnite-type, and [109.3 GPa, 170.7 GPa] for the Fe$_2$P-type TiO$_2$. The pressure interval is reduced to [165.9 GPa, 194.4 GPa] and [127.7 GPa, 152.3 GPa], in the case of $|c_p| \leq 0.2$ meV/GPa, about one order of magnitude smaller than that of SiC. More generally, for $|c_p| \leq \delta$, the pressure interval is $[P_0 + \frac{\delta - \alpha}{2\beta}, P_0 - \frac{\delta + \alpha}{2\beta}]$.

To make a comparison, the variation of the band gaps of rutile TiO$_2$ under pressure is also studied (Supplemental Material, Figure S1). The *GW* band gap (2.92 eV) of rutile at *P* = 0 compare well with the experimental value (~ 3 eV) at ambient pressure [28-30], and verifies the reliability of *GW* calculations. The fitting parameters are listed in Table I. From the data given by both GGA and *GW* calculations, the pressure coefficient of rutile turns out to be significantly larger than that of the cotunnite-type and the Fe$_2$P-type TiO$_2$. A simple comparison between the starting and ending pressure points gives that, the volume of crystal unit cell is contracted by ~ 18.3% and ~ 14.2% for the cotunnite-type and Fe$_2$P-type, with the *GW* band gaps changed by ~ +0.14 eV and ~ −0.14 eV, respectively. As a contrast, at a volume contraction of ~ 15%, the *GW* band gap of rutile TiO$_2$ is increased by ~ 0.59 eV (Figure S1). The plateau-like behavior (Figures 2(a), 2(b)) suggests that a step-like jump of the band gaps of TiO$_2$ can be observed when the applied pressure is increased gradually from 50 to 260 GPa, across the phase transition pressure point (~ 115 GPa). On the other hand, the *GW* band gaps (2.989 eV and 1.876 eV) of the cotunnite-type and Fe$_2$P-type at 160 GPa are comparable with the values (3.0 eV and 1.9 eV) estimated by previous work [33], and rationalize the empirical corrections therein.

### 3.3. Origin of the Robustness of Band Gaps upon Compression

To understand the robustness of the band gaps upon compression, we studied the pressure dependence of the valence band maximum (VBM) and conduction band minimum (CBM), the difference of which gives the value of band gap. The results by GGA and *GW* calculations are shown in Figures 2(e) and 2(f), for the cotunnite-type and $Fe_2P$-type $TiO_2$. The VBM and CBM of both phases increase almost linearly with pressure. From the *GW* (GGA) calculations, the linear rate of change ($dE/dP$) of VBM and CBM is ~ 0.0171 eV/GPa (0.0176 eV/GPa) and ~ 0.0179 eV/GPa (0.0173 eV/GPa), for the cotunnite-type $TiO_2$; and ~ 0.0139 eV/GPa (0.0144 eV/GPa) and ~ 0.0132 eV/GPa (0.0127 eV/GPa) for the $Fe_2P$-type $TiO_2$. For each high-pressure phase, the rates of change of VBM and CBM are of approximately the same value. The nearly identical rates of change explain the slow variations of band gaps with pressure. By contrast, the rates of change of VBM and CBM of rutile $TiO_2$ show considerable difference, which leads to the monotonic increase of band gap (Figure S1). As seen from Figures 2(e) and 2(f), the difference between the VBM given by GGA and *GW* calculations is very small for both phases, and the *GW* corrections on band gaps come mainly from the upshift of CBM.

*3.4. Comparison of the Structural and Electronic Properties of the Two Phases*

We have further studied the atomic and electronic properties of the two high-pressure phases. The local atomic bonding structures of the cotunnite-type and $Fe_2P$-type $TiO_2$ at $P = 115$ GPa (phase transition point), are schematically shown in the left panels of Figure 3 (a & c). The corresponding radial distribution functions (RDFs) $g_{TiO}$, which describe the spatial arrangement of O atoms around Ti, are shown in the right panels of Figure 3 (b & d). In both phases, each Ti atom is bonded with nine O atoms, i.e., nine-coordinated, which is readily deduced by doing integral within the first coordination shells as defined by the RDFs in Figures 3(b) and 3(d): 1.82 Å ≤ R ≤ 2.12 Å for the cotunnite-type and 1.82 Å ≤ R ≤ 2.06 Å for the $Fe_2P$-type. On the other hand, from their first coordination numbers, the surrounding O atoms can be classified into two types: the ones which are coordinated by four Ti atoms ($O_{4c}$) and that coordinated by five Ti atoms ($O_{5c}$), as schematically marked in Figures 3(a) and 3(c). Subjected to the stoichiometric constraints that every two O atoms are assigned to one Ti atom: $(1/4) \times n_{4c} + (1/5) \times n_{5c} = 2$ together with the condition $n_{4c} + n_{5c} = 9$, one has $n_{4c} = 4$ and $n_{5c} = 5$, which is respectively the number of four- and five-coordinated O around each Ti. In the meantime, the total number of $O_{4c}$ and $O_{5c}$ is equal in the crystal unit cell, which is 4 in cotunnite-type and 3 in the $Fe_2P$-type $TiO_2$. The nine-coordination is achieved via the periodic extension of unit cells, as indicated in Figures 3(a) and 3(c).

The sharp and discrete RDFs peaks shown in Figures 3(b) and 3(d) are the characteristics of crystalline structures [63], in which the Ti-O bond lengths ($R_{TiO}$) show a distribution instead of a single value. The averaged Ti-O bond lengths for Ti, $O_{4c}$ and $O_{5c}$ are listed in Table II. The difference of $O_{4c}$ and $O_{5c}$ is clearly reflected: the averaged value of $R_{TiO}$ is ~ 1.89 Å for the $O_{4c}$ of both phases, while the averaged $R_{TiO}$ of $O_{5c}$ is ~ 2.07 Å for the cotunnite-type and ~ 2.06 Å for the $Fe_2P$-type $TiO_2$. Moreover, judging from the averaged values of $R_{TiO}$ in Table II, the Ti atoms in the $Fe_2P$-type structure can be divided into two types: Type A ($R_{TiO}$ ~ 2.00 Å) and Type B ($R_{TiO}$ ~ 1.95 Å). The classification agrees with previous work [33].

The electron densities associated with the wave function of VBM and CBM, i.e., $|\psi_{VBM}|^2$ and $|\psi_{CBM}|^2$, are displayed in Figure 4, for the two high-pressure phases of $TiO_2$ at $P$ = 115 GPa. Clearly, the electronic states of VBM come mainly from O atoms and that of CBM come mainly from Ti atoms. Using the Bader charge analysis [64, 65], we are able to assign the number of electrons intuitively to each atom. The results are tabulated in Table II. It is interesting to find that, for both phases, different types of atoms (e.g., $O_{4c}$ vs $O_{5c}$; and Ti_A vs Ti_B in the $Fe_2P$-type) are distinguished by the different number of Bader charges. We go on to study the characteristics of VBM and CBM by projecting the corresponding wave functions on spherical harmonics. The obtained *lm*-components of VBM and CBM are summarized in Table III. For the cotunnite-type $TiO_2$, the wave function of VBM mainly consists of the $p_x$ and $p_y$ orbitals of O atoms, and the wave function of CBM comes mainly from the superposition of the $d_{z^2}$ and $d_{x^2-y^2}$ orbitals of Ti atoms. For the $Fe_2P$-type $TiO_2$, major component of the wave function of VBM is the $p_z$ orbitals of O, and the wave function of CBM involves mainly the $d_{z^2}$ orbitals of Ti, together with minor hybridization with the *s* and *p* orbitals of O.

*3.5. Understanding the Difference of Band Gaps*

As mentioned above, at pressure $P \sim 115$ GPa, the two high-pressure phases of TiO$_2$ have equal enthalpy, the same coordination numbers of Ti and O in the first coordination shells, and similar averaged value of Ti-O bond lengths (Table II). Besides, the volume per formula unit is also very close, which is $\sim 20.253$ Å$^3$/TiO$_2$ for the cotunnite-type and $\sim 20.197$ Å$^3$/TiO$_2$ for the Fe$_2$P-type. Despite these similarities, their band gaps differ by $\sim 1.06$ eV. Such difference should originate from the different crystal structures (Supplemental Table I): An orthogonal unit cell with four formula units (Z=4) for the cotunnite-type and a hexagonal unit cell with Z=3 for the Fe$_2$P-type. The positional difference of RDFs peaks (Figures 3(b), 3(d)) beyond the first coordination shell is a consequence of different O-Ti-O and Ti-O-Ti angles. This leads to different ion-electron interaction potential $V_{ext}$ and consequently the changes of the quasi-particle energies and wave functions (Table III). To the first-order perturbation of correction [58], the quasiparticle energies within the *GW* method is given by: $E_n \approx \varepsilon_n + \langle \varphi_n | \Sigma(E_n) - V_{xc} | \varphi_n \rangle$, where $\varepsilon_n$ is the eigenvalue of the corresponding Kohn-Sham orbital $\varphi_n$, and $V_{xc}$ is the exchange-correlation potential. At $P = 115$ GPa, the VBM given by GGA (*GW*) calculations for the cotunnite-type TiO$_2$ is $\sim 0.31$ eV (0.18 eV) lower than the Fe$_2$P-type. Consequently, the different band gaps are mainly attributed to the values of CBM: $E_{CBM} \approx \varepsilon_{CBM} + \langle \psi_{CBM} | \Sigma(E_n) - V_{xc} | \psi_{CBM} \rangle$. Our calculations find that, both terms $\varepsilon_{CBM}$ and $\langle \psi_{CBM} | \Sigma(E_n) - V_{xc} | \psi_{CBM} \rangle$ of the cotunnite-type are larger than that of the Fe$_2$P-type, differing by a value of $\sim 0.74$ eV for the first term and $\sim 0.15$ eV for the second. Addition of the differences between VBM and CBM results in a value of $\sim 1.07$ eV, compares well with the value (1.06 eV) obtained by solving Eq. (2).

## 4. Conclusions and outlook

In conclusion, we have studied the large-pressure-scale behavior of the structural and electronic properties of TiO$_2$, a wide band gap semiconductor. Calculation of enthalpies suggests the coexistence of two high-pressure phases across a wide range of pressures. For pressures which spanning over an interval of $\sim 150$ GPa, the band gaps ($E_g$) of both phases are rather robust to compression. The related pressure coefficients are found to be significantly smaller than that of diamond and SiC, which has the smallest pressure coefficient ever reported by experiments. For both phases of TiO$_2$, the variation of $E_g$ with pressure is well described by a quadratic polynomial. The robustness of $E_g$ is due to the nearly identical rates of changes of the VBM and CBM with pressure. Such unusual properties may have potential applications in optical devices at extreme conditions such as high pressure. Detailed analysis on crystal structures and the wave function characteristics of VBM and CBM helps to understand the difference in $E_g$ of the two structures. Finally, the pressure dependence of the optical properties, and the expected gap closure and metallization with further increasing pressures, will be the subject of a forthcoming work.

**Supplementary data**

The supplementary data related to this article can be found on the website.

**Acknowledgements**

This work is supported by the Director's Research Funds at the Institute of Solid State Physics (ISSP) of Chinese Academy of Sciences (CAS). We gratefully acknowledgment the support from the staff of the Hefei Branch of Supercomputing Center of CAS, and the crew of Center for Computational Materials Science of the Institute for Materials Research, Tohoku University for their continuous support of the SR16000 supercomputing facilities. We also thank Professor Yoshiyuki Kawazoe of Tohoku University, and Professors Zhi Zeng and Xianlong Wang of the ISSP of CAS, for their reading and helpful comments on the manuscript.

**Figures & Captions**

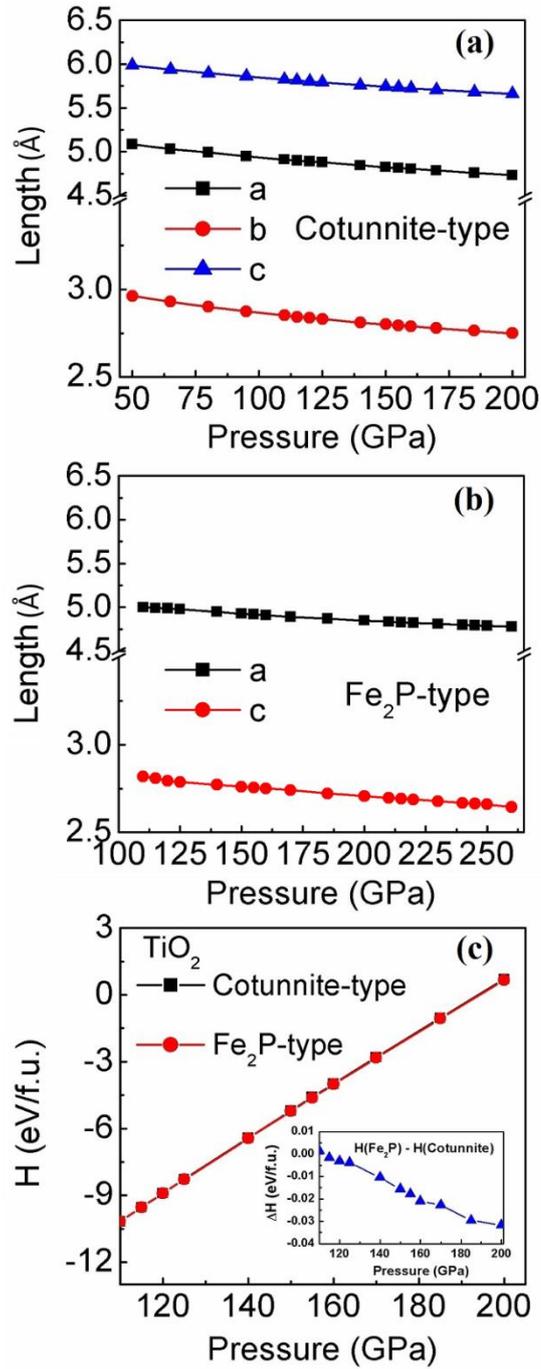

**Fig. 1.** Pressure dependence of the lattice parameters of the cotunnite-type (**a**) and Fe$_2$P-type (**b**) TiO$_2$. (**c**) The enthalpies of the two high-pressure phases of TiO$_2$, as a function of pressure. The enthalpy $H = U + PV$, where $U$ is the total energy obtained by DFT calculations, and $P$ is pressure and $V$ is volume of the system. The symbol f. u. is the abbreviation for formula units (equivalently, Ta$_2$O$_5$ unit).

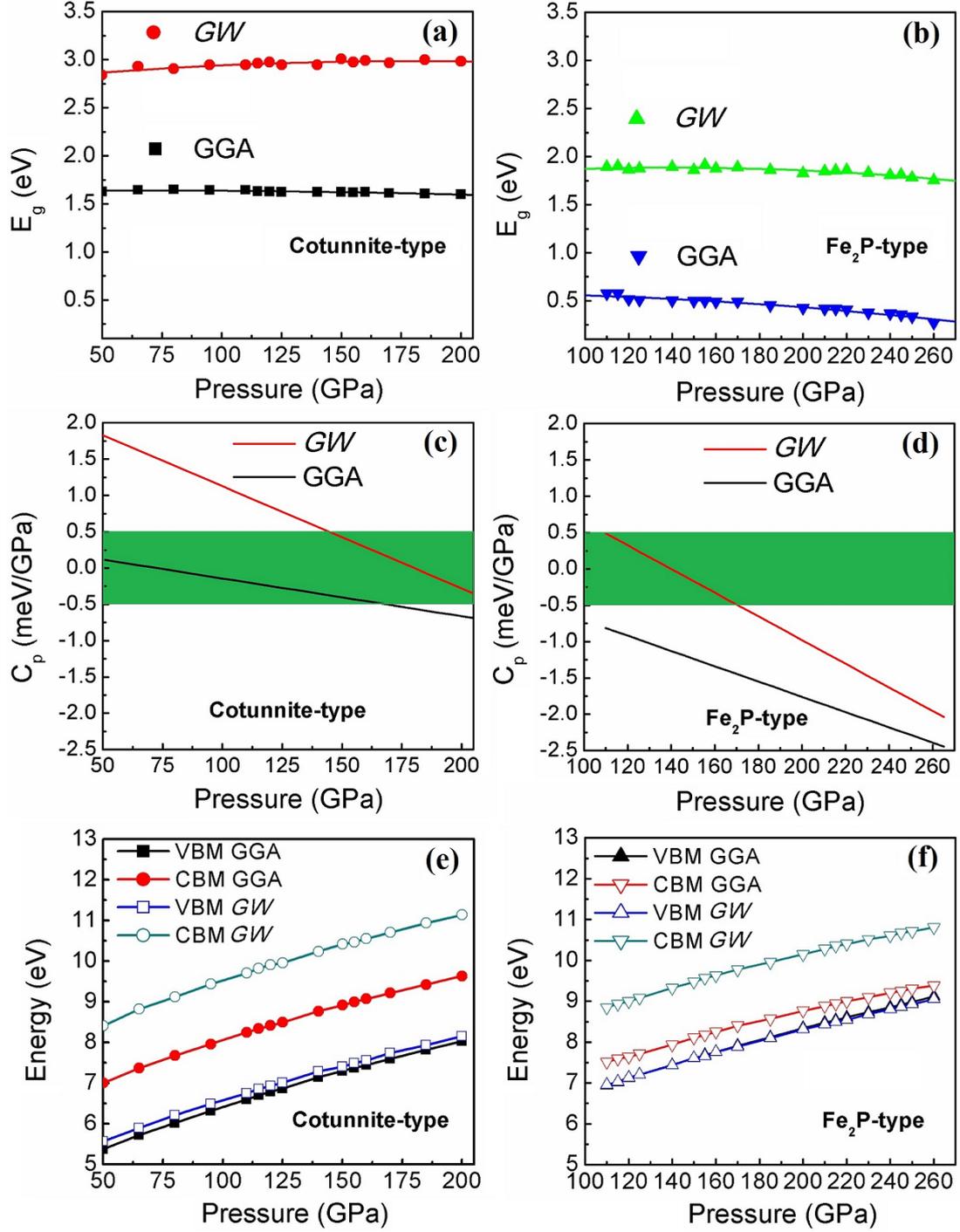

**Fig. 2.** Pressure dependence of the band gaps (Panels (**a**) and (**b**)), pressure coefficients (Panels (**c**) and (**d**), the region $|c_p| \leq 0.5$ meV/GPa is highlighted), and the valence band maximum (VBM) and conduction band minimum (CBM) (Panels (**e**) and (**f**)) of the two TiO$_2$ phases. The left panels (**a**, **c**, **e**) are for the cotunnite-type and the right panels (**b**, **d**, **f**) are for the Fe$_2$P-type. The solid lines in panels (**a**) and (**b**) represent the least-squares fits to the GGA and *GW* data points of band gaps.

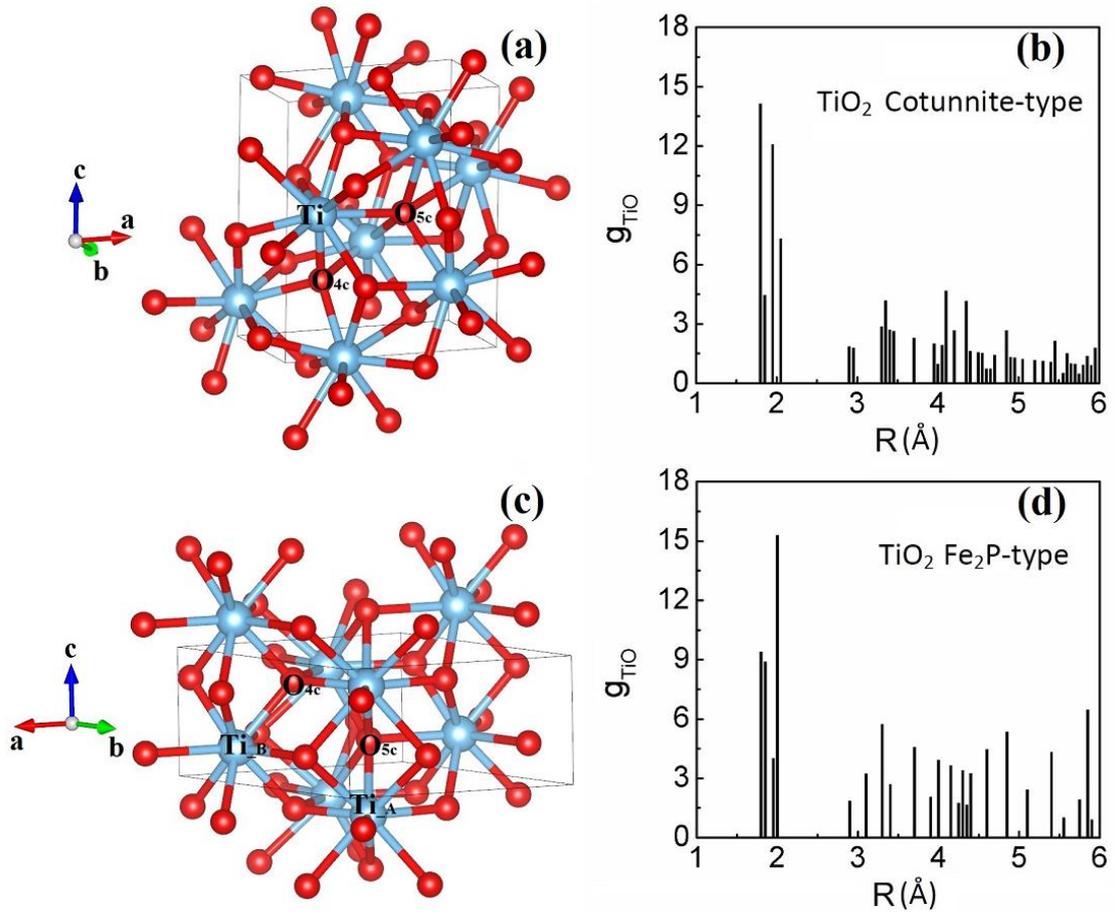

**Fig. 3.** Crystal structures and radial distribution functions (RDFs) $g_{TiO}$ for the cotunnite-type (Panels (**a**) and (**b**)) and the $Fe_2P$-type (Panels (**c**) and (**d**)) $TiO_2$, at $P = 115$ GPa. The different types of Ti and O atoms are schematically marked.

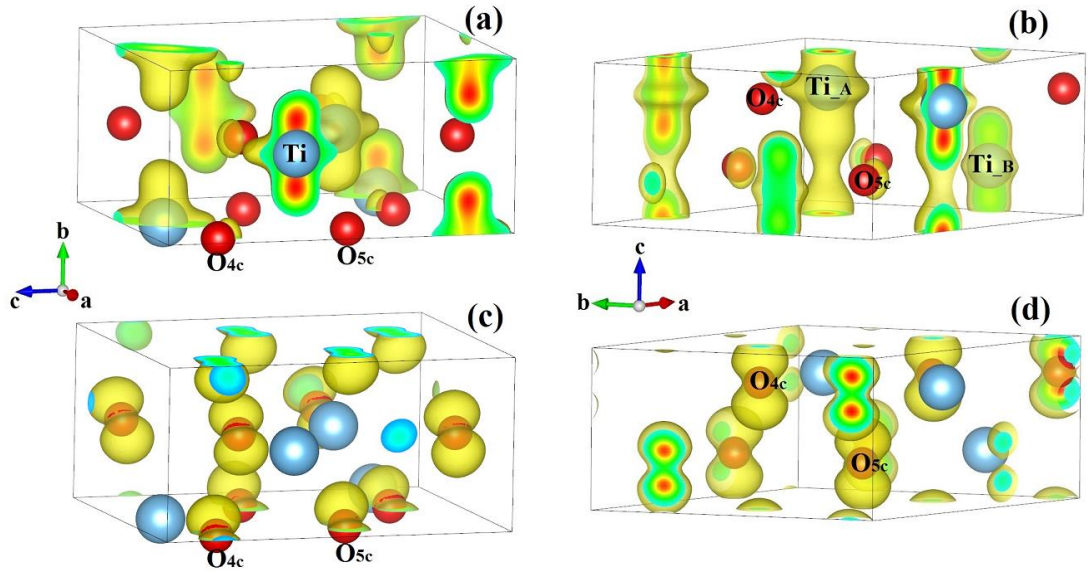

**Fig. 4.** Spatial distribution of the electron densities of the CBM ($|\psi_{CBM}|^2$, upper panels) and VBM ($|\psi_{VBM}|^2$, lower panels) of the two TiO$_2$ phases at $P$ = 115 GPa. The left panels (**a**, **c**) are for the cotunnite-type and the right panels (**b**, **d**) are for the Fe$_2$P-type.

**Table I.** The values of parameters $E_g (P_0)$, $α$, and $β$ in Eq. (4), which are obtained via the least squares method, for describing the pressure dependence of GGA and *GW* band gaps of the two high-pressure $TiO_2$ phases and rutile $TiO_2$. For the parameter $E_g (P_0)$, the calculated ($E_g (P_0)_{Cal.}$) and fitted ($E_g (P_0)_{Fit}$) values are listed for comparison.

|  | Method | $E_g (P_0)_{Cal.}$ (eV) | $E_g (P_0)_{Fit.}$ (eV) | $α$ (meV/GPa) | $β$ (meV/GPa$^2$) |
|---|---|---|---|---|---|
| $TiO_2$ Cotunnite-type | GGA | 1.626 | 1.637 | 0.117 | $-2.601 \times 10^{-3}$ |
|  | *GW* | 2.842 | 2.867 | 1.831 | $-7.035 \times 10^{-3}$ |
| $TiO_2$ $Fe_2P$-type | GGA | 0.574 | 0.548 | -0.813 | $-5.260 \times 10^{-3}$ |
|  | *GW* | 1.893 | 1.880 | 0.489 | $-8.153 \times 10^{-3}$ |
| $TiO_2$ Rutile | GGA | 1.664 | 1.653 | 4.583 | $9.989 \times 10^{-3}$ |
|  | *GW* | 2.920 (3.03)[a] | 2.914 | 12.876 | -0.035 |

[a]Experimental value [Refs. 28-30].

**Table II.** At $P$ = 115 GPa, the averaged Ti-O bond lengths ($R_{TiO}$), the first coordination numbers of the Ti (CNO) and O (CNTi) atoms, and the Bader charges associated with the valence band maximum (VBM) and the conduction band minimum (CBM), which are assigned to each type of atoms of the two $TiO_2$ phases. For both VBM and CBM, the total number of electronic states is 2 (spin-degenerate).

|  |  | $R_{TiO}$ (Å) | CNO | CNTi | Bader_VBM[a] | Bader_CBM[a] |
|---|---|---|---|---|---|---|
| $TiO_2$ Cotunnite-type | Ti | 1.987 | 9 | --- | 0.0195×4 | 0.4450×4 |
|  | $O_{4c}$ | 1.887 | --- | 4 | 0.2286×4 | 0.0427×4 |
|  | $O_{5c}$ | 2.066 | --- | 5 | 0.2502×4 | 0.0115×4 |
| $TiO_2$ $Fe_2P$-type | $Ti_A$ | 2.003 | 9 | --- | 0.0016×2 | 0.6699×2 |
|  | $Ti_B$ | 1.948 | 9 | --- | 0.0126×1 | 0.4383×1 |
|  | $O_{4c}$ | 1.893 | --- | 4 | 0.3369×3 | 0.0216×3 |
|  | $O_{5c}$ | 2.058 | --- | 5 | 0.3226×3 | 0.0517×3 |

[a]The format of Bader charges of each type of atoms: Bader charge ($e$) × the number of atoms.

**Table III.** Calculated *lm*-components of the wave function of the valence band maximum (VBM) and the conduction band minimum (CBM) of the two $TiO_2$ phases at $P = 115$ GPa.

|  |  | $s$ | $p_y$ | $p_z$ | $p_x$ | $d_{xy}$ | $d_{yz}$ | $d_{z^2}$ | $d_{xz}$ | $d_{x^2-y^2}$ |
|---|---|---|---|---|---|---|---|---|---|---|
| $TiO_2$ Cotunnite-type | VBM | 0.002 | 0.454 | 0.019 | 0.446 | 0.004 | 0.013 | 0.007 | 0.000 | 0.026 |
| | CBM | 0.067 | 0.000 | 0.033 | 0.017 | 0.000 | 0.000 | 0.232 | 0.032 | 0.537 |
| $TiO_2$ $Fe_2P$-type | VBM | 0.000 | 0.000 | 0.962 | 0.000 | 0.000 | 0.000 | 0.000 | 0.000 | 0.000 |
| | CBM | 0.064 | 0.016 | 0.000 | 0.016 | 0.000 | 0.000 | 0.820 | 0.000 | 0.000 |

**Supplementary data for "Unexpected Robustness of the Band Gaps of TiO$_2$ under High Pressures"**

Yong Yang

*Key Laboratory of Materials Physics, Institute of Solid State Physics, Chinese Academy of Sciences, Hefei 230031, China.*

1. **Pressure dependence of the band gap, VBM and CBM of rutile TiO$_2$**

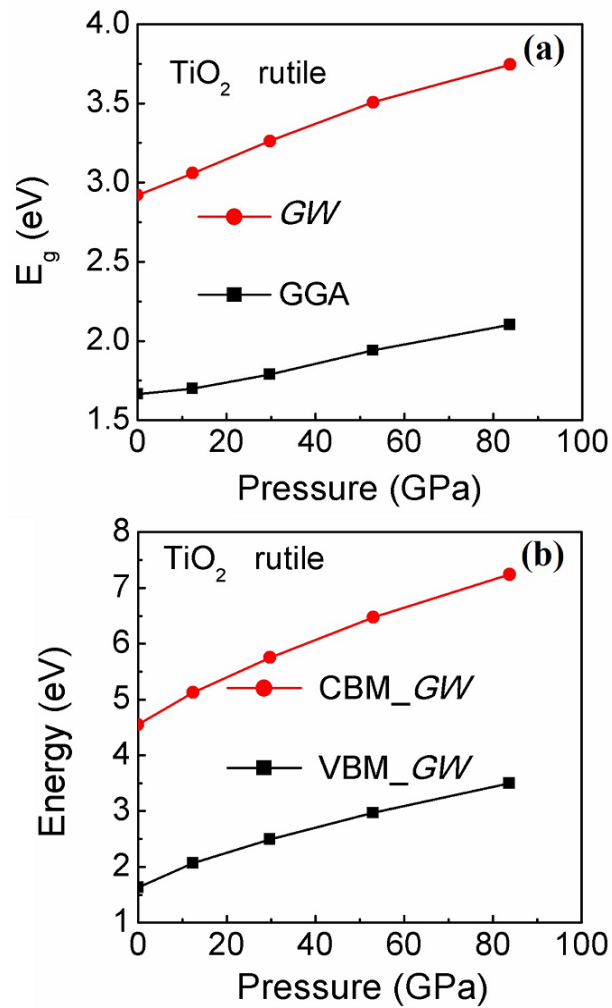

**Fig. S1.** Pressure dependence of the band gap (**a**), VBM and CBM (**b**) of rutile TiO$_2$, obtained by GGA and *GW* calculations.

2. **Supplementary Table SI.** The lattice parameters, the volume of unit cell ($V_{cell}$) at $P$ = 115 GPa; and the Murnaghan equation of state (EOS) parameters of the two high-pressure $TiO_2$ phases.

| Phase (Space group) | Lattice Parameters | | | $V_{cell}$ ($Å^3$) | EOS Parameters | | |
|---|---|---|---|---|---|---|---|
| | $a$ (Å) | $b$ (Å) | $c$ (Å) | | $B_0$ (GPa) | $B'$ | $V_0$ ($Å^3$) |
| Cotunnite-type (No. 62) | 4.900 | 2.843 | 5.815 | 81.007 | 230.6 | 4.6 | 104.845 |
| $Fe_2P$-type (No. 189) | 4.990 | 4.990 | 2.810 | 60.595 | 289.1 | 4.1 | 76.682 |